\pdfoutput=1

\documentclass[12pt,english,a4paper]{article}

\usepackage{amsmath,amsfonts,amssymb,babel,slashed,empheq,tensor}
\usepackage[utf8]{inputenc}
\usepackage{cite}
\usepackage[pdftex,breaklinks]{hyperref}
\usepackage{amsmath}
\usepackage{amsfonts}
\usepackage{array}
\usepackage{amssymb}
\usepackage{latexsym}
\usepackage{mathrsfs}
\usepackage{braket}		
\usepackage{setspace}
\usepackage{mathrsfs}
\usepackage{graphicx}
\usepackage{xcolor}
\usepackage{mathtools}
\usepackage{slashed}
\usepackage{overpic}
\usepackage[all]{xy}
\usepackage{tikz-cd}
\usepackage{slashed}
\usepackage{empheq}
\usepackage{float}
\usepackage{mathtools}
\usepackage{tikz}
\usepackage{bm}
\usepackage{pgfplots}
\pgfplotsset{compat=1.18}
\def\rlwd{.9pt}
\def\lhexbrace{\kern1pt%
\setstackgap{S}{0pt}\def\stackalignment{l}
\ThisStyle{\scalerel*{%
  \stackunder[-\rlwd]{%
    \stackon[-\rlwd]{\roundrule{\rlwd}{4pt}}{\rotatebox{60}{\roundrule{4pt}{\rlwd}}}%
  }{\rotatebox{-60}{\roundrule{4pt}{\rlwd}}}%
}{\SavedStyle[}}}
\def\rhexbrace{%
\setstackgap{S}{0pt}\def\stackalignment{r}
\ThisStyle{\scalerel*{%
  \stackunder[-\rlwd]{%
    \stackon[-\rlwd]{\roundrule{\rlwd}{4pt}}{\rotatebox{-60}{\roundrule{4pt}{\rlwd}}}%
  }{\rotatebox{60}{\roundrule{4pt}{\rlwd}}}%
}{\SavedStyle[}}\kern1pt}

\def\state#1#2{\big|\mkern-6mu
 {\scriptsize \begin{array}{c}#1\\#2\end{array}}\mkern-6mu
\big)}
\def\statex#1#2{\big({\scriptsize\mkern-6mu\begin{array}{c} #1\\#2\end{array}\mkern-6mu}\big|}

\makeatletter

\newcommand{\mso}{\mathfrak{so}}

\newcommand{\hs}{\mathfrak{hs}}
\newcommand{\msu}{\mathfrak{su}}

\newcommand{\Tr}{\text{Tr}}

\newcommand{\nn}{\nonumber}
\def\obar{\overline}

 \def\one{\mbox{1 \kern-.59em {\rm l}}}

 \def\cQ{{\cal Q}}
 \def\cM{{\cal M}}
 \def\cN{{\cal N}}
 \def\cH{{\cal H}}
 \def\cF{{\cal F}}
 \def\cC{{\cal C}
 \def\cO{{\cal O}}}
 
 \def\cA{{\cal A}}
 \def\cO{{\cal O}}

\def\TT{{\bf T}}

\newcommand{\DeltaQ}{\Delta^{\!(Q)}}

\newcommand{\End}{\mathrm{End}}
\newcommand{\tr}{\mathrm{tr}}
\newcommand{\NC}{{\rm NC}}
\newcommand{\YM}{{\rm YM}}

\def\R{{\mathbb R}} \def\C{{\mathbb C}} 

\def\a{\alpha}  \def\b{\beta}
\def\g{\gamma}

\def\t{\tau} 
\def\L{\Lambda}

\newcommand{\und}{\underline}

\newcommand{\newV}[1]{{#1}}

 \allowdisplaybreaks[3]

\makeatother

\begin{document}

\renewcommand{\title}[1]{\vspace{10mm}\noindent{\Large{\bf#1}}\vspace{8mm}}
\newcommand{\authors}[1]{\noindent{\large #1}\vspace{5mm}}
\newcommand{\address}[1]{{\itshape #1\vspace{2mm}}}

\begin{titlepage}

\begin{center}

\title{ {\Large Quantum spacetime and quantum fluctuations \\[1ex]
in the IKKT model at weak coupling} }

\vskip 10mm

\authors{Harold C.\ Steinacker}

\makeatletter{\renewcommand*{\@makefnmark}{}
\footnotetext{E-mail: \texttt{harold.steinacker@univie.ac.at}}\makeatother}

\vskip 3mm

 \address{
{\it Faculty of Physics, University of Vienna\\
Boltzmanngasse 5, 1090 Vienna, Austria  } }

\bigskip

\vskip 1.4cm

\textbf{Abstract}
\vskip 3mm

\begin{minipage}{14.8cm}%
\vskip 3mm

This paper aims to clarify conceptual aspects of emergent structure in IKKT-type matrix models. Even without any adjustable parameters in the action, non-trivial matrix vacua do acquire a meaningful coupling constant, as well as two distinct uncertainty scales: a) the scale of noncommutativity of the matrix background, and b) the scale of quantum fluctuations of the matrices under the path integral. These scales are estimated for two prototypes of matrix backgrounds, known as Moyal-Weyl quantum plane and covariant quantum spacetime.
Their relative importance separates two regimes: 1) the {\bf semi-classical regime} interpreted in terms of semi-classical noncommutative geometry, and 2) 
the {\bf deep quantum regime} usually interpreted in terms of holography.
 The quantum fluctuations \newV{around the background} are shown to be negligible in the weak coupling regime. This \newV{justifies a basic assumption in previous work on the emergent 3+1-dimensional semi-classical geometry and (quantum) gravity in suitable vacua.}

\end{minipage}

\end{center}

\end{titlepage}

\tableofcontents



\section{Introduction}

Despite the success of quantum field theory leading to the standard model of particle physics, the proper inclusion of gravity in the framework of quantum mechanics remains an enigma. At scales accessible to experiment or observation, gravity is governed by general relativity (GR), which describes how the metric of spacetime is determined by matter. However, 
it is generally expected that the classical concept of spacetime ceases to make sense below the Planck scale, where gravity becomes strongly coupled and  quantum effects become important. Moreover,
 GR is ultimately inconsistent due to inevitable singularities, such as at the Big Bang and in the center of black holes.
The situation is analogous to the classical instability of atoms,
and resolving these issues will most likely require to go beyond the classical framework of geometry and gravity.

There are many proposals how a consistent quantum theory including gravity might look like, including e.g. canonical or loop quantum gravity, asymptotic safety, holography, and string theory. This is not the place to assess them, but most researchers would agree that all approaches have issues, and there is no commonly accepted theory or framework.

This paper discusses the foundations of one candidate for such a theory: an emergent theory of (quantum) gravity defined by a distinguished matrix model known as IKKT or IIB model. This model is considered as a complete theory and can be viewed as a
non-perturbative definition of string theory, but it offers an unorthodox origin for spacetime and gravity. However
there are  different approaches even within this matrix model, leading to misconceptions and preventing a  wider adoption of this line of research.

The present paper aims to clarify these foundational issues, and to distinguish two different regimes of the matrix model: the strongly coupled regime appropriate for holography, and the weakly coupled regime leading to an emergent semi-classical matrix geometry.
This distinction does make sense even though there is no free parameter in the matrix action, as the coupling constant can be absorbed in the matrices.
The point is that the coupling constant is only defined 
{\em on a non-trivial background or vacuum}  given by some noncommutative matrix configuration.
In other words, the coupling constant arises from spontaneous symmetry breaking (SSB).
Such stable backgrounds are well-known to exist, as illustrated by D3 branes which are BPS solutions of the model. 
Note that supersymmetry of the background is not required for this mechanism, while supersymmetry of the model is crucial for several reasons.

The two regimes indicated above lead to very different concepts of emergent geometry. We focus here on the weakly coupled regime, where quantum spacetime is described through some  noncommutative algebraic structure that can be well approximated by a semi-classical symplectic manifold.
The present paper provides a consistency check and a justification for this description for two explicit 3+1-dimensional backgrounds (Moyal-Weyl quantum plane and covariant quantum spacetime), by demonstrating that the typical matrix fluctuations due to quantum effects are much smaller than the matrix background itself. 
It is quite evident that this result holds very generally for other noncommutative matrix backgrounds, in the weak coupling regime.

\newV{Specifically, we distinguish and estimate two distinct notions of uncertainty: 1) the uncertainty due to quantum fluctuations under the matrix ("path") integral, and 2) the uncertainty due to the noncommutative structure of the matrix background. The latter characterizes the quantum geometry defined by the matrix background, which a priori has nothing to do with Planck's constant $\hbar$.}

Even though the results of this paper are based on tree-level considerations in the bosonic sector of the model, supersymmetry certainly helps to carry over the results to the interacting case, at weak  coupling.
In the IKKT model, the 1-loop effective action has been worked out in recent years leading to an understanding of gravity \cite{Steinacker:2023myp,Steinacker:2021yxt,Steinacker:2024huv}, and justifying the presence of fuzzy extra dimensions \cite{Manta:2025tcl}. Since the approach is perturbative, 
this does not require a single, ''dominant'' 3+1-dimensional background; it only requires the background to be sufficiently long-lived in some sense. Remarkably, numerical simulations strongly suggest that SSB 
towards a 3+1 dimensional background geometry 
actually occurs at the non-perturbative level in the IKKT model \cite{Anagnostopoulos:2026qvz,Anagnostopoulos:2026utg,Hirasawa:2024dht,Nishimura:2019qal,Hatakeyama:2019jyw}. For other related work including Euclidean signature see e.g. \cite{Anagnostopoulos:2022dak,Nishimura:2004ts}.

The semi-classical description clearly breaks down in the strongly coupled regime, where the quantum fluctuations of all matrices dominate. This is then the domain of  holography leading to a  9+1-dimensional geometry, which is underlying much recent work on this and similar matrix models,  cf. \cite{Maldacena:1997re,Hartnoll:2024csr,Ciceri:2025wpb,Komatsu:2024bop,Asplund:2012tg,Berenstein:2008eg,Itzhaki:1998dd}. That geometry would then be described through the deep quantum regime of a strongly coupled gauge theory, which
is much more difficult to understand analytically. Making contact with physics then requires compactification, which provides rich structure \cite{Laurenzano:2025vfh} but leaves the nonperturbative realm of the matrix model.

\section{Matrix Theory: the IKKT or IIB matrix model}

In this paper, we consider Matrix Theory as defined by the IKKT or IIB matrix model
\begin{align}
 \label{IIB-action}
S_{\rm IIB}[\TT,\Psi] = 
\frac 1{g^2}\mbox{Tr}\left([\TT_{\dot a}, \TT_{\dot b}] [\TT^{\dot a}, \TT^{\dot b}]
+  \bar{\Psi}\Gamma^{\dot a}[\TT_{\dot a}, \Psi]\right) \
\end{align}
Here $\TT^a, \ a=0,...,9$ are hermitian matrices acting on a Hilbert space $\cH$, and $\Psi$ are matrix-valued Majorana-Weyl
spinors of $SO(9, 1)$. The indices are contracted with $\eta_{\dot a \dot b} = {\rm diag}(-1,1,...,1)$ with Lorentzian signature, or $\delta_{\dot a \dot b}$ in the Euclidean case.
The model is invariant under global $SO(9,1)$ or $SO(10)$ rotations and translations
\begin{subequations}
\begin{align}
      \TT^a &\to \Lambda^a_b \TT^b   \label{Lorentz}\\ 
     \TT^a &\to \TT^a + c^a \one  \label{shift}
\end{align}
\end{subequations}
(extended to fermions in the obvious way), gauge transformations 
\begin{align}
\TT^a \to U^{-1} \TT^a U, \qquad \Psi \to U^{-1} \Psi U
\end{align}
and maximal supersymmetry \cite{Ishibashi:1996xs}. 
This model is also invariant under the scale transformations
\begin{align}
 S_{\rm IIB}[\TT,\Psi;g] = S_{\rm IIB}[c\TT,c^{3/2}\Psi;c^2 g]
\end{align}
for any real $c$.
Therefore $g$ can be absorbed in the matrices, and henceforth we will set 
\begin{align}
    g=1 \ .
\end{align}
This means that there is a priori {\em no adjustable parameter} that could be used to separate weak and strong coupling regimes. These concepts only make sense in non-trivial vacua after spontaneous symmetry breaking (SSB), as discussed below.

The quantization of this matrix model is defined by integrating over the space of all matrices.
In the Lorentzian model this amounts to an oscillatory integral, which is defined through a suitable $i\varepsilon$ prescription indicated by $S_{\varepsilon}$. Then 
correlators are  defined as
\begin{align}
 \langle \cO(T)\rangle := \frac 1{Z_\varepsilon }\int dT d\Psi \cO(T) \  e^{\frac i\hbar S_\varepsilon[T]} \ , \qquad 
 Z_\varepsilon = \int dT d\Psi \  e^{\frac i\hbar S_\varepsilon[T]} 
 \label{correlators}
\end{align}
where $\cO(T)$ is some matrix observable, i.e. some function of the matrices $T^a$. 
In the Euclidean case, the exponential weight would be 
$e^{-\frac 1\hbar S[T]}$.
The subscript $\varepsilon$ will be dropped in the following, and we will set $\hbar = 1$. There are several possibilities for this $i\varepsilon$ regularization \cite{Anagnostopoulos:2026qvz,Karczmarek:2022ejn,Anagnostopoulos:2022dak,Hatakeyama:2021ake}; perhaps the simplest one is to 
perform an infinitesimal ''Wick rotation''
$\TT^0 \to e^{i\varepsilon} \TT^{0},\ \TT^i  \to  \TT^{i}$. These integrals are very difficult to compute numerically due to the oscillatory nature of the matrix integral. Nevertheless, recent progress led to strong evidence for the emergence of 3+1-dimensional backgrounds\footnote{These nonperturbative studies typically require some extra "regulator" terms such as a mass term, which may serve to focus on particular backgrounds. This is perfectly in line with the present approach.} \cite{Anagnostopoulos:2026qvz,Anagnostopoulos:2026utg,Hirasawa:2024dht,Nishimura:2019qal,Hatakeyama:2019jyw}.

\subsection{Nontrivial vacua and spacetime from SSB}

Our basic assumption or hypothesis is that 
for  large $N$, there are non-trivial matrix configurations of the model
which are stable -- or at least sufficiently metastable -- at the quantum level.
This will be indicated by
\begin{align}
\boxed{ \
 \langle \TT^{\dot a} \rangle = \obar T^{\dot a} \
 \ }
 \label{VEV-MM}
\end{align}
where $\obar T^{\dot a} \in \End(\cH)$ is some non-trivial {\bf matrix configuration}, i.e. a set of 9+1 hermitian matrices;
the bar will often be omitted. 
Such a matrix configuration will be considered as "vacuum" or {\bf background}, which can be interpreted as
spacetime in suitable cases. The matrix integral \eqref{correlators} is then evaluated as an integral of fluctuations around that background, and the background is required to be sufficiently stable in this  perturbative sense.

For matrix models with extra quadratic or cubic terms, the existence of non-trivial matrix vacua is well established. For example, a (Euclidean) action of the form 
\begin{align}
 S_{\rm YM+CS}[\TT]  = 
\mbox{Tr}\left([\TT_{\dot a}, \TT_{\dot b}] [\TT^{\dot a}, \TT^{\dot b}] + g \varepsilon_{\dot a\dot b\dot c}\TT^{\dot a}\TT^{\dot b}\TT^{\dot c} + m^2 \TT^{\dot a} \TT_{\dot a}\right)   
\end{align}
clearly
admits fuzzy sphere solutions $\obar T^{\dot a} = c J^{\dot a}$, where $J^{\dot a}$ are $\msu(2)$ generators with spin $j\gg 1$.  At the quantum level, one typically finds a nontrivial ''fuzzy sphere'' phase where \eqref{VEV-MM} holds, 
and another (''high-temperature'', matrix) phase of random matrix fluctuations around the trivial vacuum \cite{Iso:2001mg,Delgadillo-Blando:2007mqd,Delgadillo-Blando:2008cuz,Azuma:2005bj}, depending on the parameters. Nontrivial fuzzy sphere vacua also arise in the polarized IKKT model \cite{Komatsu:2024bop,Komatsu:2024ydh,Hartnoll:2024csr}.

For the IIB model, the existence of non-trivial vacua is more subtle. However in the limit $N\to\infty$, such noncommutative vacua are known to exist for the IIB model, given by the quantum plane solutions discussed in section \ref{sec:MW}. These are exact classical solutions which preserve some supersymmetry (i.e. they are BPS solutions), and are therefore protected from quantum corrections; they correspond to so-called $D3$ branes in IIB string theory. Their fluctuations are governed by noncommutative $\cN=4$ SYM theory. 
This example strongly suggests that 
in the case of large but finite $N \gg 1$, there should exist at least ''approximate'' vacua which are sufficiently long-lived or metastable to serve as spacetime. Stability 
must be checked for each candidate background by studying the fluctuations, which holds true for the BPS background in section \ref{sec:MW}, and ''essentially'' holds for the background in section \ref{sec:covar-quant}; more precisely a deformation thereof will be at least metastable in the  $N\to\infty$ limit.

The important point is that a background describing spacetime is not given a priori, but {\bf emerges} dynamically from
the matrix model; it can be viewed as a condensate in the matrix model, much like in condensed matter physics.
In such a non-trivial vacuum, the matrix d.o.f. will be parametrized as
\begin{align}
    \TT^{\dot a} = \obar T^{\dot a} + \cA^{\dot a}
    \qquad \mbox{with} \quad 
    \big\langle \cA^{\dot a}\big\rangle = 0
\end{align}
by construction.
Such a matrix configuration $\obar T^{\dot a}$ should be considered as {\bf classical}, \newV{even if it describes} a noncommutative spacetime.
This is analogous to the familiar concept of spontaneous symmetry breaking (SSB) in QFT:
in a non-trivial vacuum, some scalar fields acquire a non-trivial VEV
$\langle \phi^a \rangle = \obar\phi^a \neq 0$, and the fluctuations around this background describe the low-energy physics in that vacuum.
Quantum fluctuations in the vacuum  $\obar T^{\dot a}$  then lead to non-vanishing uncertainties of the matrix entries 
\begin{align} 
   \DeltaQ \cA^{\dot a}_{ij} := \sqrt{\Big\langle |\cA^{\dot a}_{ij}|^2\Big\rangle} 
\end{align}
in some basis of $\cH$.
One central issues is whether these quantum uncertainties $\DeltaQ \cA^{\dot a}$ are small or large compared to the (dominant matrix elements of the) background:
\begin{align}
\begin{aligned}
 \DeltaQ \cA^{\dot a}_{ij} & \ll |\obar T^{\dot a}_{ij}|  \qquad \quad \mbox{\bf semi-classical regime}  \\[1ex]
     \DeltaQ \cA^{\dot a}_{ij} & \gg |\obar T^{\dot a}_{ij}|  \qquad \quad  \mbox{\bf deep quantum regime}  
\end{aligned}
\label{class-background-crit-1}
\end{align}
This criterion requires some refinement,
since \eqref{class-background-crit-1} clearly depends on the choice of basis, and only the ''dominant'' matrix elements $\obar T^{\dot a}_{ij}$ should be considered; it is meaningless for $\obar T^{\dot a}_{ij} = 0$. 
Moreover, the diagonal matrix elements can be changed by a trivial shift $\TT^{\dot a} \to \TT^{\dot a} + c^{\dot a}\one$, and are hence not significant for the intrinsic structure of the background.
A more refined, localized characterization will be provided below.

A priori, matrix vacua might also be commutative $[\obar T^{\dot a},\obar T^{\dot b}] = 0$. This would lead to a number of issues including zero modes and a divergent action for fluctuations; see e.g. \cite{Aoki:1998vn,Ho:2025htr} for a discussion. 
Here we will focus on the generic case of noncommutative backgrounds, where these problems do not arise.

The semi-classical regime
provides the basis for an approach where spacetime emerges as matrix background $\obar T^\mu$, which encodes an effective geometry discussed below.  In contrast, holographic approaches are relevant in the deep quantum regime, where quantum fluctuations are dominant. 
Then the VEV is no longer significant, so that the effective dimension is always 9+1 as in string theory. Reducing dimensions via SSB is therefore only meaningful in the weakly coupled regime under consideration here.

\subsection{Noncommutative backgrounds and quantum fluctuations}

The naive criterion \eqref{class-background-crit-1} clearly depends on the basis, and therefore requires some refinement. 

The matrix backgrounds of interest here  describe almost-local quantum spaces. Roughly speaking this means that 
commutators of matrices are much smaller than their products, so that the matrices are mutually compatible rather than independent and random.
More precisely, the matrix elements of the commutators 
\begin{align}
  i\cF^{\dot a \dot b} :=  [\TT^{\dot a},\TT^{\dot b}]
\end{align}
are much smaller than the matrix elements of  products $\TT^{\dot a}\TT^{\dot b}$, in a suitable sense. Then the matrices can ''almost'' be simultaneously diagonalized, i.e. they have only few significant off-diagonal elements. 
Such spaces can typically be considered as quantized symplectic spaces \cite{Steinacker:2020nva}. 
More specifically,
one can \newV{define a set of optimally localized {\bf quasi-coherent} states $|t\rangle$ as ground states of a displacement operator $H_t := \sum_{\dot a} (\TT^{\dot a} - t^{\dot a}\one)^2$}.
They have minimal noncommutative uncertainty 
\begin{align}
    \Delta_{T} := \max_{\dot a}\Delta \TT^{\dot a}|_t := \max_{\dot a}\sqrt{\langle t| (\TT^{\dot a} - {\bf t}^{\dot a})
(\TT^{\dot a} - {\bf t}^{\dot a})  |t\rangle } , 
\end{align}
and satisfy 
$\TT^{\dot a}|t\rangle \approx t^{\dot a}|t\rangle$
up to corrections of order $ \Delta_{T}$.
Then the expectation values 
\begin{align}
 {\bf t}^{\dot a} :=   \langle t|\TT^{\dot a}|t\rangle \  \ \in \cM \   \subset \R^{9,1}
\end{align}
typically sweeps out some submanifold $\cM$ in target space,  denoted as ''embedded quantum space''.
This makes sense as long as the uncertainty $\Delta _T$
 is much smaller than the characteristic size of this $\cM $.
Such matrix configurations are denoted as almost-local quantum spaces \cite{Steinacker:2020nva,Steinacker:2024unq}. 
The states $|t\rangle$ are over-complete and somewhat ambiguous; they are analogs of the coherent states in quantum mechanics. In particular,
\begin{align}
    \Delta_{T} \approx \sqrt{\max_{\dot a,\dot b}
    |\langle t|\cF^{\dot a \dot b}|t\rangle|}  
\end{align}
is a (local) measure for the noncommutativity of the background, and hence for the (dominant) off-diagonal matrix elements linked to that state.
We therefore propose the following refined criterion for the semi-classical regime \eqref{class-background-crit-1}:
\newV{
\begin{align}
\label{semi-class-regime}
\boxed{ \
  \DeltaQ \cA|_t := \max_{\dot a,t'} \sqrt{\Big\langle \langle t|\cA_{\dot a} |t'\rangle \langle t'|\cA_{\dot a} |t\rangle  \Big\rangle} \ \ll \  
  \Delta _T 
\ }
\end{align}
}
The lhs measures the strength of the (matrix elements of the) quantum fluctuations in the local basis $|t\rangle$,
while the rhs characterizes the off-diagonal matrix elements\footnote{
note that diagonal matrix elements 
are not significant for the intrinsic structure of the background, and
can be changed by a trivial shift $\TT \to \TT + c\one$ of the background.} of the background. 
\newV{We allow for $t'\neq t$ on the rhs to ensure that $ \DeltaQ \cA|_t$ captures the full matrix fluctuation $\cA$, but one might also restrict to $t'=t$.}
This criterion states that the quantum fluctuations of the matrix background are small compared with the intrinsic (local) noncommutativity scale of the background geometry, which allows treating the background as a classical noncommutative space.

Although the local states $|t\rangle$ are somewhat ambiguous, the criterion is well-stated due to their over-completeness, and should hold for any almost-local bases.
We will see that this condition also implies that the quantum corrections to the metric are much smaller than the classical \newV{metric}.
It is essential here to use the local states for the estimates.

\newV{For matrix models with Minkowski signature, 
the integration measure is no longer positive. It is then not obvious how to measure the size of quantum fluctuations. Nevertheless, we propose to generalize the above criterion for the semi-classical regime as follows:
\begin{align}
\label{semi-class-regime-Mink}
\boxed{ \
  \DeltaQ \cA|_t := \max_{\dot a,\, |t'-t'|< \Delta_T} \sqrt{\Big|\Big\langle \langle t|\cA_{\dot a} |t'\rangle \langle t'|\cA_{\dot a} |t\rangle\Big\rangle\Big|} \ \ll \  
  \Delta _T 
\ }
\end{align}
Even though the lhs is less restrictive than in the Euclidean case, it still provides a useful criterion. Here we restrict to $t'\approx t$, because light-like long string modes might violate the bound;
this will be illustrated in sections \ref{sec:Mink-MW} and \ref{sec:quant-fluct-covar}.}

Note that this criterion fails for commutative backgrounds (except for the diagonal elements). This suggests that 
commutative backgrounds should not be considered as classical, because off-diagonal quantum fluctuations are never small. On noncommutative backgrounds, there are of course extra off-diagonal quantum fluctuations, but they
can be sub-leading compared with the dominant ones of the background. Hence 
only noncommutative backgrounds provide a sharp classical vacuum structure.

In the specific examples of $\R^4_\theta$ or $\cM^{3,1}_n$, one can alternatively find a discrete basis $\{|n_1,n_2\rangle\}$ or  $\{|n_1,n_2,n_3\rangle\}$ where the non-vanishing off-diagonal matrix elements are  as follows
\begin{align}
 \TT^{\dot a} |n_1,n_2\rangle \propto \sum \chi_{n_i} |n_1 \pm 1,n_2\pm 1\rangle \ + \mbox{(diag)}
\end{align}
with characteristic scale $\chi = O(\Delta _T)$. These states are analogous to the oscillator basis in quantum mechanics.
The (off-diagonal) matrix elements are comparable to those of the (quasi)coherent states, more precisely with their uncertainty $\Delta _T$. Therefore we can choose either basis and arrive at the same criterion.

The bottom line is that
there are two conceptually very different and independent types of uncertainties in nontrivial vacua of the matrix model:
\begin{enumerate}
    \item 
the quantum mechanical uncertainty $\DeltaQ$ of some observable under the matrix path integral \newV{involving an emergent $\hbar$}, and
    \item
     the intrinsic (noncommutative) uncertainty $\Delta _T$ of some given background, 
     \newV{which a priori has nothing to do with $\hbar$}. 
\end{enumerate}
The semi-classical regime \newV{of the matrix model} is characterized by 
\eqref{semi-class-regime}.
If this is violated, the effective frame and metric would receive significant contributions from quantum fluctuations, which would invalidate any semi-classical approach.

In the following
we will elaborate and illustrate these quantities for two explicit matrix backgrounds.

\section{Background 1:  Moyal-Weyl quantum plane}
\label{sec:MW}

One candidate background for the matrix mode is given by 
\begin{align}
    \TT^\mu = X^\mu
\end{align}
for $\mu=0,...,3$ or $\mu=1,...,4$ in the Minkowski or Euclidean case, respectively. 
The remaining matrices $\obar T^{\dot a}$ are zero.
Here $X^\mu$ generate the Moyal-Weyl quantum plane $\R^{4}_\theta$ defined by
 \begin{align}
   [X^\mu,X^\nu] = i\theta^{\mu\nu}\ \one \ 
   \label{MW-generic-CCR}
 \end{align}
with $\theta^{\mu\nu} = - \theta^{\nu\mu}$ a generic rank 4 antisymmetric matrix.
The $X^\mu$ are viewed as quantizations of the
Cartesian coordinate functions $x^\mu \sim X^\mu$ on $\R^{4}$ satisfying the
Poisson brackets 
\begin{align}
 \{x^\mu,x^\nu\} = \theta^{\mu\nu} = 
\begin{pmatrix} \begin{pmatrix}
                        0 & \theta_1 \\
                        -\theta_1 & 0
                       \end{pmatrix} &   \\
                        &  \begin{pmatrix}
                        0 & \theta_2 \\
                        -\theta_2 & 0
                       \end{pmatrix} 
                 \end{pmatrix}
\end{align}
in a suitable basis. For simplicity we will focus on the balanced case $\theta_1 = \theta_2$.
The scale of noncommutativity is given accordingly by
\begin{align}
\Delta^4 := 
 \sqrt{\det(\theta^{\mu\nu})} \ 
\label{L-NC-MW}
\end{align}
with dimension $L^4$.
There is a canonical (Weyl) quantization map
\begin{align}
\label{quantiz-map}
    \cQ:\quad \cC(\R^4) & \to \End(\cH) \nn\\
      e^{i k x} &\mapsto e^{i k X}
\end{align}
which is an isometry w.r.t. the inner products defined by on both sides via
\begin{align}
\int \frac{\Omega}{(2\pi)^2} \phi(x) = \Tr(\cQ(\phi)), 
\qquad \Omega = \rho_M d^4 x, \qquad \rho_M = \frac{1}{\Delta^4}
\end{align}
where $\Omega$ is the symplectic volume form on $\R^{4}_\theta$.
This structure is the same in the Euclidean or Minkowski case\footnote{Cleary $\theta^{\mu\nu}$ is not compatible with the causal structure in the Minkowki case; this issue will be remedied on covariant quantum spacetime.}, while the metric structure will be discussed below. 

A natural set of optimally localized states on $\R^4_\theta$ is given by the standard coherent states $|x\rangle$ for $x\in\R^4$, which are characterized by the coherence length 
$\Delta$: 
\begin{align}
    |\langle x|y\rangle| \sim e^{-\frac 14|x-y|_g} \ .
\end{align}
 Here
\begin{align}
    g_{\mu\nu}  \approx \frac{1}{\Delta^2}\delta_{\mu\nu}
\end{align}
(in Cartesian coordinates, assuming isotropic $\theta^{\mu\nu}$ in the last form)
 is the Euclidean quantum metric (i.e. the pull-back of the Fubini-Study metric on the space of states \cite{Steinacker:2020nva}), not to be confused with the (possibly Lorentzian) effective metric.

\subsection{Yang-Mills gauge theory on $\R^4_\theta$}
\label{sec:YM-MW}

Now consider  fluctuations $\TT^{\dot a} = \obar T^{\dot a} + \cA^{\dot a}$
on the non-trivial vacuum 
\begin{align}
\obar T^{\dot a} = 
\begin{pmatrix}
    X^\a \\ 0
\end{pmatrix} 
\end{align}
corresponding to $\R^4_\theta$.
We want to understand the classical dynamics of the tangential fluctuations $\cA^\mu$ on $\R^{4}_\theta$ and for the transversal  fluctuations $\phi^i$.
\newV{These matrix fluctuation modes are identified with functions on space(time) via a quantization map \eqref{quantiz-map}, which provides the relation with field theory in the path integral approach.}
To understand the matrix model action for these fluctuations, we rewrite the tangential and transversal fluctuations as 
\begin{align}
\cA^{\dot a} = 
\begin{pmatrix}
    \cA^\a \\ \cA^i
\end{pmatrix} = \begin{pmatrix}
   \theta^{\a\mu} A_\mu \  \\  \phi^i
\end{pmatrix} .
\label{A-tang-trans}
\end{align}
One then finds using $[X^\a,.] = i\theta^{\a\mu}\partial_\mu$
\begin{align}
[\TT^\a,\phi] &= [X^\a + \cA^\a,\phi] = i \theta^{\a\mu}
(\partial_\mu \phi - i [A_\mu,\phi]) \, =: \,  i \theta^{\a\mu} D_\mu \phi, \nn\\
\,[\TT^\a,\TT^\b] &= i\theta^{\a\mu} \theta^{\b\nu}
(-\theta^{-1}_{\mu\nu} + F_{\mu\nu})
\label{XX-gauge}
\end{align}
where $F_{\mu\nu} = \partial_\mu A_\nu - \partial_\nu A_\mu - i [A_\mu,A_\nu]\,$ is the
$U(1)$ field strength on $\R^4_\theta$.
Then the action takes the following semi-classical form 
\begin{align}
S[A,\phi] 
 &= \int\limits_{\R^{4}_\theta} \frac{d^{4} x}{(2\pi)^2}\,\rho_M\Big(
-\g^{\mu\mu'}\, \g^{\nu\nu'}\,F_{\mu\nu}\,F_{\mu'\nu'}
\, - \g^{\mu\nu} \delta_{\mu\nu}  
 - 2 \g^{\mu\nu}\, D_\mu\phi^i D_\nu \phi_i
 +  [\phi^i,\phi^j][\phi_i,\phi_j]\Big)   \nn\\
&= \int\limits_{\R^{4}_\theta} \frac{d^{4} x}{(2\pi)^2}\,\sqrt{|G|}
\Big(- \frac{1}{g_\YM^2} G^{\mu\mu'}\, G^{\nu\nu'}\,F_{\mu\nu}\,F_{\mu'\nu'}
\, - G^{\mu\nu} \delta_{\mu\nu}  \nn\\
& \qquad\qquad
 - 2 G^{\mu\nu}\, D_\mu\phi^i D_\nu \phi_i
 + {g_\YM^2} [\phi^i,\phi^j][\phi_i,\phi_j] \Big) 
\label{action-YM-scalars-R2n}
\end{align}
(dropping surface terms) for the effective metric 
\begin{align}
     G^{\mu\nu} &= \rho^{-2}\gamma^{\mu\nu}, \qquad
     \gamma^{\mu\nu} = \eta_{\a\b}\theta^{\a\mu}\theta^{\b\nu} 
\end{align}
and $\eta_{\a\b} \to \delta_{\a\b}$ in the Euclidean case. The metric is unimodular in the Cartesian coordinates 
\begin{align}
 \sqrt{|G_{\mu\nu}|} &=  1 \ .
\end{align}
The dilaton
\begin{align}
    \rho^2 &= \frac{\sqrt{|G_{\mu\nu}|}}{\rho_M} = \Delta^4
\end{align}
transforms as a scalar, and defines the effective coupling
constant\footnote{\newV{Restoring units and  $g$, this should be replaced by $g^2_{\YM}=\frac{g^2 \hbar}{\Delta^4}$, with $\Delta$ considered as a number in Cartesian matrix coordinates $x^\mu \sim X^\mu$.}}
\begin{align}
    g^2_\YM = \frac{1}{\rho^2} \  = \frac{1}{\Delta^4} 
    \label{YM-coupling-MW}
\end{align}
which governs the gauge fields as well as the scalar fields. 
The gauge symmetry $\TT^\a \to U^{-1} \TT^\a U$ 
acts on the fluctuations as expected:
\begin{align}
A_\mu \to U^{-1} A_\mu U + i U^{-1} \partial_\mu U, \qquad
\phi^i \to U^{-1} \phi^i U \ .
\label{gauge-A-phi}
\end{align}
We therefore recover the well-known fact that the matrix models reduces to noncommutative $U(1)$ Yang-Mills gauge theory on $\R^4_\theta$. This extends\footnote{Superficially, the $U(1)$ sector appears to be weakly coupled because all fields transform in the adjoint, and hence almost commute. However this is not true for the UV sector of the theory, which is given by the string modes discussed below. Then $g_\YM \ll 1$ ensures that even the string modes are weakly coupled.} straightforwardly to 
$U(n)$ upon replacing the background by a stack of coinciding $\R^4_\theta$ ''branes'', 
\begin{align}
\label{stack-branes}
\obar T^{\dot a} = 
\begin{pmatrix}
    X^\a\otimes \one_n \\ 0
\end{pmatrix} 
\end{align}
and $g_\YM$ is recognized as Yang-Mills coupling. The above background can in fact be interpreted as (a stack of) D-branes in string theory, equipped with effective metric $G_{\mu\nu}$ corresponding to the open string metric in some decoupling limit \cite{Seiberg:1999vs}, as well as a 2-form given by $\theta^{-1}_{\mu\nu}$. 

We note the following message: even though the matrix model does not have a coupling constant, 
in a non-trivial vacuum a dimensionless coupling arises which may be small or large, depending on the background.

\subsection{Quantum fluctuations on $\R^4_\theta$}
\label{sec:quant-fluct-MW}

Now  we wish to estimate the strength of the quantum fluctuations 
$\cA$ on Moyal-Weyl space. This is measured by the matrix correlators
\begin{align}
     \Big\langle \cA^\a \cA^\b \Big\rangle =  
     \theta^{\a\mu} \theta^{\b\nu} \Big\langle A_{\mu}  \cA_{\nu} \Big\rangle \ 
\end{align}
for the free matrix action.
We impose for simplicity the Lorentz gauge\footnote{this is equivalent to the Lorentz gauge $\partial^\mu A_\mu = 0$ using the effective metric $G_{\mu\nu}$.}
$[X^\mu,\cA_\mu] = 0$ using the Feynman gauge parameter, e.g. by adding a Faddeev-Popov ghost term or BRST ghosts; see e.g. \cite{Steinacker:2024unq} for details.
Expanding the fluctuations of the gauge field in plane waves 
\begin{align}
    A_\mu = \int d^4 k\; A_\mu(k) e^{i k X}
\end{align}
and noting that
\begin{align}
    \Box e^{i k X} = \rho^2 G^{\a\b}k_\a k_\b e^{i k X} 
\end{align}
where $\Box = [T^\a,[T_\a,.]]$,
the free action \eqref{action-YM-scalars-R2n} for the gauge fields then takes the form 
\begin{align}
    S_0[A]
    &=  \frac{(2\pi)^2}{g_\YM^2}
    \int d^4 k A_\mu(-k) (G^{\a\b}k_\a k_\b +i \varepsilon) A^\mu(k) \ .
\end{align}
We first consider the Euclidean case. Then the propagator in momentum space is the standard one,
\begin{align}
    \Big\langle A_\mu(k)  A_\nu(k') \Big\rangle = 
     \frac{g_\YM^2 }{(2\pi)^2} \,\frac{\delta^{(4)}(k-k') }{ |k|^2_G }G_{\mu\nu} \ .
\end{align}
Reinserting the plane waves $e^{i k X}$  as elements of $\End(\cH)$, we obtain the matrix 2-point function
\begin{align}
    \Big\langle \langle i|A_\mu|j\rangle \langle k|A_\nu|l\rangle \Big\rangle 
    = \frac{g_\YM^2 }{(2\pi)^2}
    \int d^4 k \frac{\langle i|e^{i k X} |j\rangle \langle k| e^{-i k X}|l\rangle}{|k|^2_G }G_{\mu\nu} \ .
\end{align}
The desired matrix elements in the localized states $|x\rangle$ are obtained 
using
\begin{align}
    e^{i k X}|x\rangle = |x + \tilde k\rangle, \qquad 
    \tilde k^\mu = k_\nu \theta^{\nu\mu}
\end{align}
and
\begin{align}
\label{plane-wave-expect-string}
    \langle x'|e^{i k X}|x\rangle = \langle x'|x + \tilde k\rangle \sim e^{-\frac 14|x'-x-\tilde k|^2_g} \ 
\end{align}
(dropping a phase).
Therefore the local matrix elements of quantum fluctuations of $A$ are estimated by
\begin{align}
   \Big\langle |\langle x'|A_\mu|x\rangle|^2  \Big\rangle 
    &= \frac{g_\YM^2 }{(2\pi)^2}
    \int d^4 k \frac{|\langle x'|e^{i k X} |x\rangle|^2 }{|k|^2_G }   
    \sim \frac{g_\YM^2 }{(2\pi)^2}
    \int d^4 k \frac{e^{-\frac 12|x'- x- \tilde k|^2_\delta /\Delta^2} }{|k|^2_G }
    \label{fluc-square-1}
\end{align}
recalling $|.|^2_g = 
|.|_\delta^2/\Delta^2$.
The $k$ integral clearly has an effective UV cutoff set by 
$\L_\NC = \frac{1}{\Delta}$. 
Hence for  $|x-x'|_\delta \ll \Delta$, we can approximate $|x'- x-\tilde k|^2_\delta \approx |\tilde k|^2_\delta$, so that
\begin{align}
   \Big\langle |\langle x'|A_\mu|x\rangle|^2  \Big\rangle 
&\approx \frac{g_\YM^2 }{(2\pi)^2}
    \int d^4 k \frac{1}{|k|^2_G } e^{-\frac 12|k|^2_\delta \Delta^2} \
 \approx \frac{g_\YM^2 }{2\Delta^2} , \qquad 
 |x-x'|_\delta \ll \Delta
 \label{fluc-square-1.5}
\end{align}
recalling $|G_{\mu\nu}| = 1$.
Note that this is {\bf finite} for $x\to x'$ without further regularization.
In the non-local regime $ |x-x'|_\delta > \Delta$, we can evaluate \eqref{fluc-square-1} using 
\begin{align}
  e^{-\frac 12|x'- x- \tilde k|_\delta^2 /\Delta^2}
\approx \frac{(2\pi)^2}{\Delta^4} \delta^{(4)}(\theta^{-1}(x'- x)- k), 
\end{align}
which gives 
\begin{align}
   \Big\langle |\langle x'|A_\mu|x\rangle|^2  \Big\rangle 
 &\approx 
   \frac{g_\YM^2 }{\Delta^4}
     \frac{1}{|\theta^{-1}(x - x')|^2_G}
 \  \approx \ g_\YM^2  \frac{1}{|x-x'|_\delta^2} , \qquad \mbox{for}\quad
  |x-x'|_\delta > \Delta \ .
    \label{fluc-square-2}
\end{align}
Hence
the non-local quantum fluctuations $ \big\langle |\langle x'|A_\mu|x\rangle|^2  \big\rangle$ corresponding to long string modes extend over large distances, which is a typical UV/IR effect of noncommutative field theories.
In terms of the original matrix fluctuations $\cA^\a = \theta^{\a\nu} A_\nu$, we obtain the estimate
\begin{align}
     \Big\langle |\langle x'|\cA_\a|x\rangle|^2  \Big\rangle 
 \ \leq \ \frac{g_\YM^2 }{2} \Delta^2 .
  \label{fluc-square-1.7}
\end{align}
There is a shortcut to obtain this result directly from the matrix action using string modes  
\begin{align}
    \cA^{x,y} = |x\rangle\langle y| \equiv \state{x}{y}
\end{align}
\cite{Steinacker:2022kji,Steinacker:2016nsc}
which satisfy
\begin{align}
    \Box \state{x}{y} &\approx \Big((x-y)_\delta^2 + 2 \Delta _T^2\Big)\state{x}{y} , \qquad [\TT^\a,.] \state{x}{y}  \approx (x-y)^\a\state{x}{y} \nn\\
    \statex{x'}{y'} \state{x}{y} &\approx 
     e^{-(x'-x|_g^2/4}  e^{-|y'-y|_g^2/4}
\end{align}
(dropping to phase factor in the 2nd line) up to corrections set by $\Delta_T \equiv \Delta$.
The extra term $2\Delta _T^2$ arises from the noncommutativity of the matrix background \cite{Steinacker:2022kji,Steinacker:2016nsc}.
Since the string modes are approximate eigenstates of $\Box$ linking two points $x$ and $y$, they can be interpreted in terms of open strings\footnote{Note that string modes are UV modes, with energy $\geq \Delta_T$. The semi-classical low-energy modes are recovered from Gaussian packets of string modes with $|x-y| < \Delta$, which correspond to wavepackets with momentum $k =\tilde x - \tilde y$, cf. \cite{Steinacker:2024unq} chapter 6.2.} or dipoles \cite{Bigatti:1999iz}.
Hence the 2-point function for string modes
$\langle x'|\cA_\a|x\rangle = \tr(\cA_\a \state{x}{y})$
 is given by the string propagator 
\begin{align}
     \Big\langle \cA^{y,x}_\a \cA^{x,y}_\b \Big\rangle &= \left(^x_{y} \right| \frac {\delta_{\a\b} }{\Box } \left|^x_{y} \right) 
 \approx \frac{\delta_{\a\b} }{(x-y)_\delta^2 + 2\Delta _T^2 }
 \label{fluc-square-comb}
\end{align}
in agreement with both \eqref{fluc-square-2} and  \eqref{fluc-square-1.7} since $\Delta^2_T = \Delta^2 = g_\YM^{-1}$.

In particular, we conclude that the quantum fluctuations are indeed small
and the criterion \eqref{semi-class-regime} for the semi-classical regime is satisfied. Hence 
\newV{
\begin{align}
\boxed{\ 
 g_\YM  \ll  1 \qquad \Rightarrow \qquad
    \DeltaQ \langle x'|\cA_\a|x\rangle 
    = O\Big(g_\YM  \Delta _T\Big) \ \ll  \ 
    \Delta _T 
    \ }
     \label{fluc-square-comb-2}
\end{align}
}
This is precisely the condition for weak coupling  \eqref{semi-class-regime} on the given background.
We conclude that
{\bf the quantum fluctuations $\cA$ are much smaller than the background in the weak coupling regime}. In particular, 4-dimensional vacua can emerge from the model without compactification, noting that there are no modes propagating in transverse dimension\footnote{From the string theory point of view, this regime should correspond to the Seiberg-Witten decoupling limit \cite{Seiberg:1999vs}; however there are no physical 9+1-dimensional closed string modes in the weakly-coupled matrix model. Their effect only arises as induced interactions, cf. \cite{Ishibashi:1996xs,Chepelev:1997av,Steinacker:2016nsc}.}.

Incidentally,  the integral \eqref{fluc-square-1.5} in the correlation functions would diverge in 2 dimensions $\R^2_\theta$ in the IR. This seems to rule out 
semi-classical 2-dimensional vacua in the matrix model, in close analogy to the Mermin-Wagner theorem.

For completeness we also consider the (analog of the) standard 2-point function,
in order to compare with standard results. Reinserting the phase in
\eqref{plane-wave-expect-string} and \eqref{fluc-square-1} for local expectation values
\begin{align}
    \langle x|e^{i k X}|x\rangle 
    \sim e^{i k x} e^{-\frac 14|\tilde k|^2_g} \ 
\end{align}
(cf. (3.1.117) in \cite{Steinacker:2024unq}), we obtain the 2-point function
\begin{align}
     \Big\langle \langle x|A_\mu|x\rangle \langle y|A_\mu|y\rangle \Big\rangle   
     &=  \frac{g_\YM^2 }{(2\pi)^2}
    \int d^4 k \frac{e^{i k (x-y)} e^{-\frac 12|\tilde k|^2_g}}{|k|^2_G }
    \label{local-corr-R4}
\end{align}
which features an effective cutoff at $|\tilde k|_g^2 \sim  \Delta^2 |k|^2 = O(1)$  for these {\em local} UV modes. This is not a fundamental UV cutoff, rather it arises because the extreme UV modes are non-local string modes whose local expectation value vanishes. 
Again we should distinguish two regimes.
If $|x-y| \gg \Delta$, then $k < \L_\NC = \frac 1{\Delta}$ dominates the integral, where $|\tilde k|_g^2 \ll 1$ and the UV truncation is immaterial. 
We then recover the standard 2-point function
\begin{align}
   \Big\langle \langle x|A_\mu|x\rangle  \langle y|A_\mu|y\rangle \Big\rangle 
    &\sim
     \frac{g_\YM^2 }{(2\pi)^2} \frac 1{[x-y|_G^2} \ .
    \label{fluc-square-class}
\end{align}
On the other hand if $|x- y| < \Delta$, 
we recover the regularized coincidence limit \eqref{fluc-square-1.5}
\begin{align}
 \Big\langle \langle x|A_\mu|x\rangle  \langle y|A_\mu|y\rangle \Big\rangle 
     &\approx \frac{g_\YM^2 }{2 \Delta^2}
    \qquad \mbox{for}\quad |x-y| \ll \Delta \ .
\end{align}
Since $[\cF^{\a\b},.] = 0$, the one-loop contributions 
in the IKKT model
only enter at $O(\cA^4)$  due to maximal SUSY. Therefore the results for the free action also hold at one loop, and are expected to be reliable at weak coupling.

Analogous results of hold for the fluctuations of the (transversal) scalar matrix fluctuations $\phi^i$.
Of course there is no VEV in transversal directions hence the fluctuations dominate in these directions, but they are smaller than  the {\em non-trivial} matrix background.

\paragraph{Relation to oscillator basis.}

Another natural basis for the Hilbert space underlying $\R^4_\theta$
is given by the harmonic oscillator states $\{|n_1,n_2\rangle, n_i = 0,1,2,... \}$ generated by $a_1^\dagger \sim X^1 + i X^2, \ a_2^\dagger = X^3 + i X^4$. The ground state $|0,0\rangle \cong |x=0\rangle$ then coincides with the coherent state at the origin, and any other coherent state $|x\rangle = U_x |0\rangle$ can be translated to the origin using a unitary transformation $U_x$ (a similar statement holds for covariant quantum spacetime using space-like symmetries). Therefore we can assume that $x=0$, and use the estimate \eqref{fluc-square-comb}
\begin{align}
   \DeltaQ\langle 0|\cA_\a|y\rangle \leq \frac{1}{\sqrt{2\Delta_T^2 + |y|^2}} . 
\end{align}
Since the first excited oscillator state(s) are captured by states $|y\rangle$ localized at a distance of order $\Delta$ from the origin,
the above result implies an estimate for the quantum fluctuations of matrix elements  in the oscillator basis 
\begin{align}
\DeltaQ \langle 0,0| \cA^{\a} |n_1,n_2\rangle 
 = O\Big(\frac{1}{\Delta_T}\Big)
\ll 
 \langle 0,0| \TT^{\a} |n_1,n_2\rangle  = O(\Delta_T) \qquad n_{1,2} = O(1)
\end{align}
provided $g_\YM = \Delta_T^{-2} \ll 1$.
For  $n_i \gg 1$, these matrix elements correspond to the non-local sector of the theory where the quantum fluctuations are bounded by \eqref{fluc-square-comb}, while the matrix elements of the background vanish.
Finally, matrix elements where both $n_i$ and $n_i'$ are large
\begin{align}
 \langle n_1 \pm 1,n_2\pm 1| \TT^{\a} |n_1,n_2\rangle  = O(\sqrt{n} \Delta_T)
\end{align}
correspond to states localized far from the origin. Due to the translational symmetry implemented by unitary transformations $U_x$, there is no need to derive separate estimates for the corresponding quantum fluctuations.

\subsection{Minkowski signature and a possible pathology for $\R^{3,1}_\theta$}
\label{sec:Mink-MW}

If the embedding $\R^{3,1}\subset \R^{9,1}$ has Minkowski signature, then
the propagator is given by
\begin{align}
    \Big\langle A_\mu(k)  A_\nu(k') \Big\rangle =  -i
     \frac{g_\YM^2 }{(2\pi)^2} \,\frac{\delta^{(4)}(k-k') }{ |k|^2_G - i \varepsilon}G_{\mu\nu} \ .
\end{align}
Following the steps of the Euclidean case using the $i\varepsilon$ regularization in the propagator,  we can estimate the local matrix elements of quantum fluctuations $A$ as
\begin{align}
   \Big\langle |\langle x'|A_\mu|x\rangle|^2  \Big\rangle 
    &= -i\frac{g_\YM^2 }{(2\pi)^2}
    \int d^4 k \frac{|\langle x'|e^{i k X} |x\rangle|^2 }{|k|^2_G - i \varepsilon}   \
    \sim -i \frac{g_\YM^2 }{(2\pi)^2}
    \int d^4 k \frac{e^{-\frac 12|x'- x- \tilde k|^2_\delta /\Delta^2} }{|k|^2_G - i \varepsilon} \ .
    \label{fluc-square-1-Mink}
\end{align}
In the local regime $|x-x'|_\delta \ll \Delta$,
we can again approximate $|x'- x-\tilde k|^2_\delta \approx |\tilde k|^2_\delta$, so that
\begin{align}
   \Big\langle |\langle x'|A_\mu|x\rangle|^2  \Big\rangle 
&\approx -i\frac{g_\YM^2 }{(2\pi)^2}
    \int d^4 k \frac{1}{|k|^2_G - i \varepsilon} e^{-\frac 12|k|^2_\delta \Delta^2} 
 \approx (1-i)\frac{g_\YM^2 }{4\Delta^2} , \qquad 
 |x-x'|_\delta \ll \Delta
 \label{fluc-square-1.5-Mink}
\end{align}
 using 
\begin{align}
    \int d^4 k \frac{e^{- k^\mu  k^\nu \delta_{\mu\nu}/2}}{k^\mu k^\nu \eta_{\mu\nu} - i \varepsilon} 
 &= 4\pi \int\limits_{-\infty}^\infty d k_0 \int\limits_0^\infty dk k^2 \frac{e^{- (k^2 + k_0^2)/2}}{k^2 - k_0^2 - i \varepsilon} 
 = \pi^2 (1+i)
 \label{integral-formula}
\end{align}
for $\varepsilon \searrow 0$.
\newV{The result is no longer real, because the weight in the matrix integral \eqref{correlators} is complex, with an $i\varepsilon$ prescription. Nevertheless, the absolute value provides some measure for the 
size of the matrix fluctuations  $\cA^\a = \theta^{\a\nu}A_\nu$, and essentially recovers
the Euclidean estimate \eqref{fluc-square-1.5}. We therefore propose to 
use the absolute value of the fluctuations $\Big|\Big\langle |\langle x'|A_\mu|x\rangle|^2  \Big\rangle \Big|$ in the criterion \eqref{semi-class-regime-Mink} for the semi-classical regime of the matrix model.  }

In the non-local regime $|x-x'|_\delta \gg \Delta$, one finds
\begin{align}
   \Big\langle |\langle x'|A_\mu|x\rangle|^2  \Big\rangle 
 &\approx 
  -i \frac{g_\YM^2 }{\Delta^4}
     \frac{1}{|\theta^{-1}(x - x')|^2_G - i \varepsilon}
   \  \approx \ \frac{ -i g_\YM^2}{(x-x')_\eta^2 - i \varepsilon} , \qquad \mbox{for}\quad
  |x-x'|_\delta > \Delta \ .
    \label{fluc-square-2-Mink}
\end{align}
Hence
the non-local quantum fluctuations $ \big\langle |\langle x'|A_\mu|x\rangle|^2  \big\rangle$ corresponding to long string modes extend over large distances notably for light-like $|x-x'|_\eta$.
Note that the result is imaginary, reflecting the fact that string modes are in the extreme UV regime.
The shortcut using string modes works again with  
\begin{align}
    \Box \state{x}{y} &\approx \Big((x-y)_\eta^2 + 2 \Delta _T^2\Big)\state{x}{y} \ .
\end{align}
Now the string modes are approximate eigenstates of the d'Alembertian linking two points,
and the 2-point function for string modes
$\langle x'|\cA_\a|x\rangle = \tr(\cA_\a \state{x}{y})$
 is obtained immediately as 
\begin{align}
     \Big\langle \cA^{y,x}_\a \cA^{x,y}_\b \Big\rangle &= \left(^x_{y} \right| \frac {-i\eta_{\a\b} }{\Box - i \varepsilon} \left|^x_{y} \right) 
 \approx \frac{-i\eta_{\a\b} }{(x-y)_\eta^2 + 2\Delta _T^2 - i \varepsilon}
 \label{fluc-square-comb-M}
\end{align}
\newV{lhs fixed, flipping one  $ \cA^{x,y}_\a \to \cA^{y,x}_\a$}
in agreement with  \eqref{fluc-square-2-Mink} since $g_\YM = \Delta^{-2}$.

We conclude again that the quantum fluctuations are small compared with the background at weak coupling $g_\YM \ll 1$, and the criterion \eqref{fluc-square-comb-2} holds.
In particular, 3+1 dimensional physics can emerge from the model without compactification.
However, Moyal-Weyl backgrounds lead to a strange and probably pathological feature in Minkowski signature.
Although the induced target space metric $\eta_{\mu\nu}$ and the effective metric $G_{\mu\nu}$ have Minkowski signature, their causality structure and light cones are incompatible. This is reflected in two distinct types of (approximate) poles in the propagators on $\R^{3,1}_\theta$: the correlators \eqref{fluc-square-class} (extended to Minkowski case) for local matrix elements $\langle x|\cA_\a|x\rangle$ have a pole at $|x-y|_G = 0$ as expected, but the 2-point function for the non-local fluctuations  $\cA \sim |x\rangle\langle y|$ \eqref{fluc-square-comb-M}
features a new (approximate) pole at $|x-y|_\eta =0$. This 
exhibits the stringy nature of matrix models, with  string modes behaving as open strings which are on-shell if $|x-y|_\eta \approx 0$.

These distinct causality structures for $\R^{3,1}_\theta$ lead to a problem: light-like directions for the target space can be space-like from the effective metric point of view. This leads to acausal (e.g. instantaneous) long-distance interactions mediated by open string (modes) at one loop, which indicates that such a background is physically unacceptable and probably inconsistent or  unstable; cf. \cite{Gomis:2000zz} for a somewhat related discussion. In the maximally supersymmetric IKKT model, such 1-loop interactions are strongly suppressed, but the effect is still present via induced non-local interactions of the form 
$\frac{1}{|x-y|_\eta^8}$ \cite{Steinacker:2016nsc,Chepelev:1997av,Ishibashi:1996xs}. This recovers essentially the interactions mediated by massless closed strings  IIB supergravity, which in the weakly coupled matrix background 
manifest themselves as extra short-range interactions\footnote{note that the weakly coupled matrix model does not contain any explicit closed string modes, only the interactions are visible and may be interpreted in this way.}.

This problem does not arise on the covariant quantum spacetime considered below, where $G_{\mu\nu} \sim \eta_{\mu\nu}$ coincide by a conformal factor.
Then the 1-loop interaction simply amounts to a weak, short-range causal extra interaction $\sim \frac{1}{|x-y|_\eta^8}$  on spacetime, which indicates interesting novel physics that might be detectable in suitable situations.

\section{Background 2: (minimal) covariant quantum spacetime  $\cM^{3,1}$}
\label{sec:covar-quant}

Our second candidate background for the matrix model is given by 
\begin{align}
    \TT^\a = T^\a := r^{-1} M^{\a 4} , \qquad \a =0,...,3 \ .
\end{align}
Here $M^{ab}, \ a,b=0,...,5$ are generators in the doubleton representation $\cH_n$  of $\mso(4,2)$; for simplicity we consider only the minimal case $n=0$ here. 
The remaining matrices $\obar T^{\dot a}$ are zero.
Note that $r$ is a {\em number}, which sets a length scale on the background.
The algebra of operators $\End(\cH) \cong \cC(\cM)$ is a quantization of the space of functions on the symplectic space $\cM = \C P^{1,2}$, which is an $S^2$ bundle over 
spacetime $\cM^{3,1}$. We describe it through 4+4 functions or generators $T^\mu$ and $X^\mu$ subject to two constraints, which arise from the Lie algebra representation. These are
\begin{align}
  X^\mu &= r M^{\mu 5} \sim x^\mu, \qquad 
   T^\mu = r^{-1} M^{\mu 4} \sim t^\mu, \qquad \mu = 0,..,3 
    \label{Cartes-coords}
\end{align}
and they satisfy the constraint\footnote{We state the constrain in the semi-classical limit; for the precise form see e.g. \cite{Manta:2025inq}.}
\begin{align}
\label{constraints-M31}
    x^\mu t_\mu = 0, \qquad r^2 t_\mu t^\mu = - r^{-2} x_\mu x^\mu .
\end{align}
The $x^\mu$ will be used as Cartesian coordinate functions on spacetime $\cM^{3,1}$, while the $t^\mu$ describe the $S^2$ fiber over $\cM^{3,1}$.
It is convenient to also consider $X^4 = r M^{45} \sim x^4$ such that
\begin{align}
  x_a x^a &= r^2 \one, \qquad a=0,...,4 \ \nn\\
   x_4 &= r \sinh(\tau) \sim \frac{r}{2} e^\tau  
   \label{x4-explicit}
\end{align}
so that $r^2 t_\mu t^\mu = - r^{-2} x_\mu x^\mu \sim \sinh^2(\tau)$.
Here $\tau$ will play the role of a cosmic time parameter.
The algebra of functions thus
decomposes into a tower of spin $s$ valued functions on spacetime
\begin{align}
\label{hs-functions-decomp}
    \cC = \C[[x^\mu,t_\nu]]/_\sim \
    = & \ \  \cC^0 \ \oplus \ \cC^1 \ \oplus \ \cC^2
 \oplus ...  \nn\\
 & \ \ \cC^s \ni \phi^{(s)} = \phi_{\und{\mu}}(x) u^{\und{\mu}}
\end{align}
(here $\und{\mu}$ is a multi-index)
in terms of irreducible polynomials in $u^\mu = \frac{r}{\sinh\tau}\, t^\mu$ of degree $s$; here $u_\mu u^\mu = 1$ generate a space-like internal sphere $S^2$.
One can again define quantization maps compatible with $SO(4,2)$, which locally look like
\begin{align}
\label{quant-map-covar}
    \cQ:\quad L^2(\cM^{3,1} \times S^2) & \to \End(\cH) \nn\\
      e^{i k_\mu x^\mu} Y^{sm}(u) &\mapsto e^{i k_\mu X^\mu} \hat Y^{sm}
\end{align}
providing a one-to-one correspondence of classical functions on $\C P^{1,2} \cong \cM^{3,1} \times S^2$ with operators, at least in the IR regime. 
The scale of noncommutativity can be obtained from the bracket relations 
\begin{subequations}
    \begin{align}
  \{t^\mu,x^\nu\} \sim -i [T^\mu,X^\nu] &= \frac{1}{r} X^4\eta^{\mu\nu} \ \sim  \sinh(\tau)\eta^{\mu\nu}  \label{t-x-CR}, \\
    \{x^\mu,x^\nu\} \sim  -i [X^\mu ,X^\nu] &= - r^2M^{\mu\nu} 
    \sim  - \frac{r^2}{\sinh(\tau)}(t^\mu x^\nu - t^\nu x^\mu)    \,,  \\
  \{t^\mu,t^\nu\} \sim \ -i [T^\mu ,T^\nu] &= \frac{1}{r^2} M^{\mu\nu} \ \ \sim  \frac 1{r^2 \sinh(\tau)} (t^\mu x^\nu - t^\nu x^\mu) 
\end{align}
\label{T-X-CR}
\end{subequations}
together with the constraints \eqref{constraints-M31} and \eqref{x4-explicit}, evaluated at the reference point 
$x^\mu = (x^0,0,0,0)$ with $x_0 \approx x_4 = \frac r2 e^\t$ for $\tau \gg 1$. One finds
\begin{align}
\Delta_X^2 \approx r^2 \Delta^2, \qquad \Delta _T^2 = r^{-2}\Delta^2, 
\qquad \Delta^2 := \sinh(\tau) \ .
\label{L-NC-covar}
\end{align}
The trace is related to the integral w.r.t. the symplectic volume form as in \eqref{L-NC-MW},
\begin{align}
 \Tr(\cQ(\phi)) &= 
\int \frac{\Omega}{(2\pi)^3} \phi(x) 
= \int\limits_{\cM^{3,1}\times S^2} \frac{d^4x d^2 t}{(2\pi)^3 r^{2}\Delta^6} \phi
    = \int\limits_{\cM^{3,1}}\frac{d^4x}{(2\pi)^2} \rho_M [\phi]_0, 
 \nn\\
 \rho_M &= \frac{2}{r^3 x_4} .
\label{trace-covar}
\end{align}
using $\int_{S_t^2} d^2 t = 4\pi r^{-4}x_4^2$,
 where $[\Phi]_0$ denotes the average over $S^2$ normalized as 
 \begin{align}
      [1]_0 = 1 \ .
      \label{average-normaliz}
 \end{align}
Here $\rho_M$ is obtained by integrating the symplectic volume form $\Omega$ over $S^2$.

It is also possible to write down the $T^\mu$ and $X^\nu$ matrices in a discrete weight basis $|\und{i}\rangle$ using e.g. an oscillator construction of $\cH_0$, see Appendix A.5 in  \cite{Manta:2025inq}. Choosing some reference point with 
$X^0 = m r$ and minimal $X^i \approx 0$, one finds that 
some of the $X^\mu$ and $T^\mu$ (i.e. $X^0$ and $T^3$, for example)
are large $O(r^{\pm 1}m)$, while all other non-vanishing elements are $O(r^{\pm 1}\sqrt{m})$. The large matrix elements just reflect the large expectation values $\langle X^\mu\rangle, \langle T^\nu\rangle = O(r^{\pm 1} m)$ for the corresponding optimally localized state $|\xi\rangle$, and can be removed by a shift of $X^\mu$ and/or $T^\nu$. In this sense,
the intrinsic matrix elements are almost-diagonal
\begin{align}
\langle \und{i}|  T^\mu |\und{i} \pm 1\rangle = O(r^{-1}\sqrt{m}) = O(\Delta_T), \quad 
  \langle \und{i}| X^\mu |\und{i} \pm 1\rangle = O(r\sqrt{m}) = O(\Delta_X)
\end{align}
and are captured 
consistently by the scale of noncommutativity $\Delta$.

As for the Moyal-Weyl quantum plane, one can find for any point $\xi = (x^\mu,t^\nu) \in \cM$ ''quasi-coherent'' states $|\xi\rangle$
which are optimally localized  on $\cM$,
with expectation values 
\begin{align}
\label{quasi-coh-states-covar}
    \langle X^\mu,T^\nu\rangle_\xi = \xi = (x^\mu,t^\nu)
\end{align}
and uncertainties $\Delta T^\mu, \Delta X^\mu$ consistent with \eqref{L-NC-covar}. They provide again an over-complete basis, see \cite{Manta:2025inq} for an explicit construction of such states.

\subsection{Yang-Mills gauge theory on $\cM^{3,1}$}

Now consider the non-trivial vacuum\footnote{Strictly speaking, this background is only a solution in the presence of a mass term in the model, which we assume in this section for simplicity. Without such a mass term, the background would acquire an extra dependence on the cosmic time, but this does not change the essential conclusions here.}
\begin{align}
\obar T^{\dot a}  = 
\begin{pmatrix}
    T^\a \\ 0
\end{pmatrix} 
\end{align}
corresponding to $\cM^{3,1}$, and add
  fluctuations $\TT^{\dot a} = \obar T^{\dot a} + \cA^{\dot a}$
to this background.
We rewrite the tangential and transversal fluctuations as 
\begin{align}
\cA^{\dot a} = 
\begin{pmatrix}
    \cA^\a \\ \cA^i
\end{pmatrix} = \begin{pmatrix}
   e^{\a\mu} A_\mu \  \\  \phi^i
\end{pmatrix} .
\label{A-tang-trans-M31}
\end{align}
Due to the bracket relations \eqref{T-X-CR},
the matrix background $\TT^\a = T^\a \sim t^\a$ naturally act as momentum generators on covariant quantum spacetime $\cM^{3,1}$. In particular, it defines in the semi-classical regime a frame 
\begin{align}
\label{eff-frame-cov}
    e^{\a\mu} = \{t^\a,x^\mu\} = \sinh(\tau) \eta^{\a\mu}
     = \Delta^2 \eta^{\a\mu} \ .
\end{align}
One then finds using\footnote{For simplicity we only write down here the semi-classical regime.} $[T^\a,.] \sim i\{t^\a,.\} = ie^{\a\mu}\partial_\mu$
\begin{align}
[\TT^\a,\phi] &= [X^\a + \cA^\a,\phi] \sim i e^{\a\mu}
(\partial_\mu \phi - i [A_\mu,\phi]) \, =: \,  i e^{\a\mu} D_\mu \phi, \nn\\
\,[\TT^\a,\TT^\b] &\sim ie^{\a\mu} e^{\b\nu}
(B_{\mu\nu} + F_{\mu\nu})
\label{XX-gauge-M31}
\end{align}
where 
\begin{align}
    F_{\mu\nu} = \partial_\mu A_\nu - \partial_\nu A_\mu - i [A_\mu,A_\nu]\, + ...
\end{align}
 is the standard field strength on $\cM^{3,1}$, and the ellipses indicate extra terms that are negligible in the flat (i.e. late-time, here) regime. 
\begin{align}
    B_{\mu\nu} &= \frac{1}{r^2\sinh^2(\t)} M_{\mu\nu} 
    = \frac{1}{r^2\sinh^3(\t)} (t_\mu x_\nu - t_\nu x_\mu), \nn\\
    [B_{\mu\nu}]_{S^2} &= 0
\end{align}
can be viewed as a $\hs$-valued background 2-form field, which vanishes upon averaging over the internal $S^2$.
Then the semi-classical action takes again the standard Yang-Mills form
\begin{align}
S[A,\phi] 
 &\sim \int\limits_{\cM^{3,1}} \frac{d^{4} x}{(2\pi)^2}\,\rho_M\Big[
-\g^{\mu\mu'}\, \g^{\nu\nu'}\,F_{\mu\nu}\,F_{\mu'\nu'}
\, - \g^{\mu\nu} \eta_{\mu\nu}  
 - 2 \g^{\mu\nu}\, D_\mu\phi^i D_\nu \phi_i
 +  [\phi^i,\phi^j][\phi_i,\phi_j]\Big]_0   \nn\\
&= \int\limits_{\cM^{3,1}} \frac{d^{4} x}{(2\pi)^2}\,\sqrt{|G|}
\Big[- \frac{1}{g_\YM^2} G^{\mu\mu'}\, G^{\nu\nu'}\,F_{\mu\nu}\,F_{\mu'\nu'}
\, - \frac{1}{g_\YM^2} G^{\mu\mu'}\, G^{\nu\nu'}\,B_{\mu\nu}\,B_{\mu'\nu'} \nn\\
& \qquad
 - 2 G^{\mu\nu}\, D_\mu\phi^i D_\nu \phi_i
 + {g_\YM^2} [\phi^i,\phi^j][\phi_i,\phi_j] \Big]_0 
\label{action-YM-scalars-covar}
\end{align}
where $[.]_0$ denotes averaging over $S^2$ as in \eqref{trace-covar},
for the effective metric and dilaton
\cite{Steinacker:2010rh,Sperling:2019xar,Steinacker:2024unq}
\begin{align}
G_{\mu\nu} = \rho^2 \gamma_{\mu\nu} &= 2 r^{-4} \sinh(\tau) \eta_{\mu\nu}  
=  2r^{-4} \Delta^2 \eta_{\mu\nu} ,    \nn\\
 \gamma^{\mu\nu} &= \eta_{\a\b} e^{\a \mu}e^{\b \nu} = \sinh^2(\tau)\eta^{\mu\nu} 
 \label{eff-metric-def-covar}
\end{align}
in Cartesian coordinates $x^\mu$, and the dilaton $\rho$ is a scalar field given by 
\begin{align}
 \rho^2 &=  \frac{\sqrt{|G|}}{\rho_M}
 =\rho_M \sqrt{|\gamma^{\mu\nu}|}
 = \rho_M \det(e^{\a\mu})\ .
  \label{dilaton-def-covar}
\end{align}
The effective coupling constant is determined as\footnote{\newV{Note that $r$ should be considered as a number in matrix coordinates $x^\mu$. Then the standard dimensions of field theory are recovered with dimensionless $g_\YM$.}}
\begin{align}
    g^2_\YM = \frac{1}{\rho^2} \ =  \frac {r^4}{2\sinh^3(\tau)}  =  \frac {r^4}{2\Delta^6} 
    \label{YM-coupling-covar}
\end{align}
(cf. \eqref{YM-coupling-MW})
which governs the gauge fields as well as the scalar fields, and becomes weak at late times\footnote{This behavior can change for more general time-dependent backgrounds with the same structure \cite{Manta:2025tcl}.} $\tau\to \infty$. 
Here the background action is 
\begin{align}
    \frac{1}{g_\YM^2} G^{\mu\mu'}\, G^{\nu\nu'}\,B_{\mu\nu}\,B_{\mu'\nu'}
 &= - \frac {1}{\sinh(\tau)} \ .
\end{align}
We also dropped a linear term $\sim\int B F$, which is an artifact that would disappear upon stabilizing the background e.g. by adding a mass term (or an induced vacuum energy).

Even though this action looks like a Yang-Mills action, the field content is now much richer: expanding the fluctuations $\cA^a$ 
according to \eqref{hs-functions-decomp} as
\begin{align}
     \cA^{\a} = \cA^{\a\und{\mu}}(x) u_{\und{\mu}}
\end{align}
they describe a tower of higher-spin ($\hs$) gauge fields, and similarly $\phi^i = \phi^{i;\und{\mu}}(x) u_{\und{\mu}}$.
Therefore the action \eqref{action-YM-scalars-R2n} describes a  $\hs$ extension of $U(1)$ Yang-Mills theory, which is actually a gauge theory of geometry because the $\cC^1$ fluctuations describe deformations of the geometry.
Similarly on a background $T^\a \otimes \one_n$  analogous to \eqref{stack-branes}, the action describes a $\hs$ extension of nonabelian $U(n)$ Yang-Mills theory, with coupling strength $g_\YM$.
In contrast to $\R^{3,1}_\theta$, the realization of this $\cM^{3,1}$ in string theory is not known; see \cite{Castelino:1997rv} for a related discussion.

\subsection{Quantum fluctuations on $\cM^{3,1}$}
\label{sec:quant-fluct-covar}

We now wish to compute the quantum fluctuations of the background 
\begin{align}
     \Big\langle \cA^\a \cA^\b \Big\rangle =  
     e^{\a\mu} e^{\b\nu} \Big\langle A_{\mu}  \cA_{\nu} \Big\rangle \ 
\end{align}
along the lines of section \ref{sec:quant-fluct-MW}, by computing the 2-point function of the above theory at tree level. This could proceed by first fixing the (Lorentz) gauge and expanding the fields in a basis of $\hs$-valued eigenmodes $\Box \Upsilon_{lm}^{sq} = \lambda \Upsilon_{lm}^{sq}$, which is then easily inverted to give the propagator. There is a slight technical complication since the background is not flat: the (scalar) eigenmodes $\Upsilon_{lm}^{sq}$ are not strictly plane waves but more naturally organized in terms of $SO(3,1) \subset SO(4,1)$ eigenmodes. This can be done exactly 
(see \cite{Battista:2022hqn} for a detailed computation of the propagator and
\cite{Sperling:2019xar,Steinacker:2019awe} for a more group-theoretic organization), but is hardly worth the effort for our purpose. 
Since we are interested in the local fluctuations of the background (local w.r.t. the basis of localized states $|\xi\rangle$ \eqref{quasi-coh-states-covar}), it is quite clear that the details of the global geometry will not be important, and the eigenmodes $\Upsilon$ locally reduce to plane waves as in \eqref{quant-map-covar} times the $\hs$ modes $Y^{sm}$.
We can then directly carry over the computations in section \ref{sec:quant-fluct-MW} for 
$A_\mu = \sum\int d^4k A_\mu^{sm}(k) e^{i k X} Y^{sm}$, defining ''reduced'' matrices 
\begin{align}
   A_\mu^{sm}  = 
 [A_\mu Y^{sm}]_0 = \int d^4k A_\mu^{sm}(k) e^{i k X}
   \label{reduced-hs-modes}
\end{align}
which picks out a single $X$-dependent $\hs$ mode; we could restrict ourselves e.g. to the sector $s=m=0$ corresponding to ordinary spacetime modes.
These sectors are mutually orthogonal, and 
the free action \eqref{action-YM-scalars-covar} for the gauge fields then takes the form 
\begin{align}
    S_0[A]
    &=  \frac{(2\pi)^2}{g_\YM^2}
    \int d^4 k \sum_{sm} A^{sm}_\mu(-k) (\eta^{\a\b}k_\a k_\b -i \varepsilon) A^{s-m}_{\nu}(k) \eta^{\mu\nu}
\end{align}
noting that $\sqrt{|G|} G^{\mu\mu'}\, G^{\nu\nu'} = \eta^{\mu\mu'}\eta^{\nu\nu'}$.
Hence the propagator in momentum space is the standard one,
\begin{align}
\label{A-correlators-M31}
    \Big\langle A^{sm}_\mu(k)  A^{s'm'}_\nu(k') \Big\rangle =  -i
     \frac{g_\YM^2 }{(2\pi)^2} \frac{\delta^{(4)}(k-k') }{ |k|^2_\eta - i \varepsilon} \delta_{ss'} \delta_{m,-m'}\eta_{\mu\nu}
\end{align}
and the fluctuations of the modes are obtained using \eqref{reduced-hs-modes} as
\begin{align}
   \Big\langle |\langle x'|A^{sm}_\mu|x\rangle|^2  \Big\rangle 
    &= -i\frac{g_\YM^2 }{(2\pi)^2}
    \int d^4 k \frac{|\langle x'|e^{i k X}|x\rangle|^2 }{|k|^2_\eta - i \varepsilon}  \ .
   \label{fluc-square-1-covar}
\end{align}
We are mainly interested in nearly-local matrix elements with $|x - x'|_\delta < \Delta_X = r \Delta$; for the non-local ones see \eqref{fluc-square-comb-M31-string}. 
Then $\langle x'|e^{i k X}|x\rangle \approx 1$ for $k \leq (r\Delta)^{-1}$, while for larger $k$ we can estimate 
\begin{align}
    |\langle x'|e^{i k X}|x\rangle|^2 \leq e^{-\frac 12|k|_\delta^2 r^2 \Delta^2} \ .
    \label{sandwich-est-covar}
\end{align}
Hence the uncertainty scale $\Delta$ serves again as a cutoff for the otherwise UV divergent integral over $k$, and we can use \eqref{sandwich-est-covar} for all $k$.
Then
\begin{align}
   \Big\langle |\langle x'|A^{sm}_\mu|x\rangle|^2  \Big\rangle 
    &\approx -i \frac{g_\YM^2 }{(2\pi)^2}
    \int d^4 k \frac{e^{-\frac 12|k|^2_\delta r^2 \Delta^2} }{|k|^2_\eta - i \varepsilon}
    = \frac{g_\YM^2}{2r^2\Delta^2}(1-i)
 \end{align}
 using \eqref{integral-formula}.
Therefore 
\begin{align}
   \Big\langle |\langle x'|\cA^{sm}_\a|x\rangle|^2  \Big\rangle 
    &\approx (1-i)\frac{g_\YM^2  \Delta^2}{2r^2} \ 
    \qquad \mbox{for} \ \ |x - x'|_\delta < \Delta_X .
       \label{fluc-square-covar-1.5-M31}
 \end{align}    
Now recall that e.g. $\cA_\a^{s=1}$ corresponds to a fluctuation of the background $T^\a = t^\a$, which has uncertainty $\Delta_T = r^{-1} \Delta$. \newV{Taking absolute value according to \eqref{semi-class-regime-Mink}, we conclude that}
 the quantum fluctuations of the $\hs$ modes $\cA_\a^{sm}$ are again smaller than the noncommutative uncertainty  at weak coupling, and the criterion \eqref{fluc-square-comb-2} holds:
\newV{
\begin{align}
\label{weak-coupling-uncertaint<-Mink}
g_\YM \ll 1 \qquad \Rightarrow \qquad
     \Delta^{Q} \cA \approx \frac{\Delta}{\sqrt{2}r}g_\YM \ll \Delta_T \ .
\end{align}
}
A similar estimate applies to the fluctuations of the transversal scalar fluctuations $\phi^i$. 

It is worth to discuss again the shortcut using  an over-complete basis of optimally localized string modes $\cA_{\xi,\zeta} = |\xi\rangle\langle \zeta| \equiv \state{\xi}{\zeta}$,
where $|\xi\rangle, |\zeta\rangle$ are optimally localized states on $\cM \cong \C P^{1,2}$; these will be related to a specific matrix basis for the underlying representation of $SO(4,2)$ below.
The string modes are again approximate eigenstates of the matrix d'Alembertian
\begin{align}
    \Box \state{\xi}{\zeta} &\approx \Big((t_\xi-t_{\zeta})_\eta^2 + 2 \Delta_T^2\Big)\state{\xi}{\xi'}  \nn\\
    \statex{\xi'}{\zeta'} \state{\xi}{\zeta} &\approx 
     e^{-|\xi-\xi'|_g^2/4}  e^{-|\zeta-{\zeta'}|_g^2/4}
\end{align}
(up to phase).
Hence the 2-point function for string modes
$\langle \xi'|\cA_\a|\xi\rangle$
modes is given by the string propagator 
\begin{align}
     \Big\langle \cA^{\xi,\zeta}_\a \cA^{\xi,\zeta}_\b \Big\rangle &= \left(^\xi_{\zeta} \right| \frac {-i\eta_{\a\b} }{\Box - i \varepsilon} \left|^\xi_{\zeta} \right) 
 \approx \frac{-i\eta_{\a\b} }{|t_\xi-t_\zeta|_\eta^2 + 2\Delta _T^2 - i \varepsilon} \ .
 \label{fluc-square-comb-M31-string}
\end{align}
We can use this to obtain the local 2-point function for the $S^2$ harmonics
\begin{align}
\langle x|\cA^{sm}_\a|x\rangle
= \langle x|[\cA Y^{sm}]_0 |x\rangle 
= \frac 1c\int_{S^2} dt Y^{sm}(t)\langle xt| \cA_\a|xt\rangle
\end{align}
where 
\begin{align}
    c = \int_{S^2} dt' Y^{sm}(t) Y^{sm}(t) \sim 4\pi t\cdot t \sim  4\pi r^{-2}\Delta^4 
\end{align}
(the normalization is such that $Y^{00} = [Y^{00}]_0 = 1$, cf. \eqref{average-normaliz}).
Therefore
\begin{align}
      \Big\langle \langle x|\cA^{sm}_\a|x\rangle \langle x|\cA^{sm}_\b |x\rangle\Big\rangle 
     &= \frac 1{c^2} \int_{S^2} dt  \int_{S^2} dt' Y^{sm}(t) Y^{sm}(t') \Big\langle \langle xt| \cA_\a|xt\rangle
     \langle xt'| \cA_\b|xt'\rangle\Big\rangle \nn\\
     &\approx \frac {-i}{c^2} \int_{S^2} dt  \int_{S^2} dt' \frac{Y^{sm}(t) Y^{sm}(t') }{|t-t'|_\eta^2 + 2\Delta _T^2 }\, \eta_{\a\b} \ .
\end{align}
In the late-time regime, the squared radius of $S^2$ is $t\cdot t \sim r^{-2}\Delta^4 \gg r^{-2}\Delta^2 = \Delta_T^2$.
Moreover, the separation $t-t'$ is space-like at equal times.
Therefore we can approximate  
\newV{
\begin{align}
\label{prop-string-M31}
  \frac{1}{|t-t'|_\eta^2 + 2\Delta _T^2 } \approx \pi \ln(1+2r^2\Delta_T^2) \; \delta^{(2)}(t-t')  \,
\end{align}
and hence
 \begin{align}
      \Big\langle |\langle x|\cA^{sm}_\a|x\rangle|^2 \Big\rangle 
     &\approx \frac {-i}{c^2} \int_{S^2} dt Y^{sm}(t) Y^{s-m}(t) 
      = \frac {-i\pi \tau }c \sim \frac{-i}{4} \frac{r^2 \tau}{\Delta^{4}}  
\end{align}
}
which together with \eqref{YM-coupling-covar} recovers \eqref{fluc-square-covar-1.5-M31} \newV{up to a factor of order $\ln(r^2\Delta_T^2)\approx \tau$. Similar logarithmic correction factors for the propagator in the string basis were found in \cite{Steinacker:2022kji}, although their origin is not entirely clear. 
We still take it as a confirmation of the the basic result \eqref{weak-coupling-uncertaint<-Mink}.}

\section{Further topics}

\paragraph{Frame fluctuations.}

An immediate application of the above result concerns quantum fluctuations of the effective frame, which arise from quantum fluctuations  as \eqref{eff-frame-cov}
\begin{align}
    \label{eff-frame-cov-fluct}
   e^{\a\mu} = \{t^\a + \cA^\a,x^\mu\} = \obar e^{\a\mu} + \delta_\cA e^{\a\mu}
\end{align}
Since we have shown that $\cA \ll t^\a$ for typical quantum fluctuations at weak coupling, it follows that quantum fluctuations of the frame are negligible compared with the background. 
The spectral distribution of $\cA$ can be inferred from \eqref{A-correlators-M31}. Since $\delta_\cA e^{\a\mu} =  \{\cA^\a,x^\mu\}$ is a derivative of $\cA$, we obtain 
\begin{align}
    \Big\langle \delta_\cA e^{\a\mu}(k) \delta_\cA e^{\a\mu}(k') \Big\rangle \sim \frac{k k'\delta^{(4)}(k-k') }{ |k|^2_\eta - i \varepsilon} \ .
\end{align}
This is different from canonical quantum gravity, because it arises from the Yang-Mills-type action rather than the Einstein-Hilbert action. Taking into account
the induced Einstein-Hilbert action at one loop \cite{Steinacker:2023myp} would significantly modify this result and bring it more in line with the standard picture (except at very long wavelengths). Nevertheless, there should be interesting new physics in these quantum fluctuations.

\paragraph{Remark on the BFSS model.}

Finally, it is natural to ask analogous questions for the BFSS model \cite{Banks:1996vh,deWit:1988wri}. That model can be viewed as matrix quantum mechanics, and -- in contrast to the IKKT model -- it does have a well-defined coupling strength $g$ or 't Hooft coupling  $\lambda =g^2N$. Therefore there is an a priori notion of weak or strong coupling, and one may ask about phase transitions at finite temperature. 
Much work in the literature is focused on the strong-coupling (or low temperature) regime, which provides a holographic relation to IIA string theory in 9+1 dimensions and gauge/gravity duality \cite{Berkowitz:2016jlq,Asplund:2012tg,Berenstein:2008eg,Itzhaki:1998dd}, as well as a conjectural relation to (or definition of) M theory \cite{Banks:1996vh}.

Nevertheless, one may consider the possibility of  long-lived non-trivial states with reduced symmetry, which might be interpreted as spacetime branes, cf. \cite{Brahma:2022ifx,Brahma:2021tkh}. 
This would be interesting in the semi-classical, weakly coupled regime on such a background. 
However, there is no compelling candidate for such a background describing 3+1--dimensional spacetime
(although there are certainly many solutions of the classical matrix equations of motion, cf. \cite{Kabat:1997im,Arnlind:2004br}).
For this and other reasons, the IKKT model offers a more natural path towards 3+1-dimensional physics.
Nevertheless, the BFSS point of view may provide useful insights to the IKKT model; for example, the high-temperature limit of the BFSS model reduces to the bosonic part of the Euclidean IKKT model \cite{Kawahara:2007ib}, as explored in \cite{Brahma:2022ifx}.

\section*{Discussion and outlook}

The main point of this paper is a disambiguation of distinct approaches to large-$N$ Yang-Mills matrix models, and to justify the 
semi-classical approach based on non-trivial matrix backgrounds at weak coupling. The basic observation is that a well-defined notion of coupling strength arises on non-trivial backgrounds, even though the underlying matrix action has no meaningful coupling constant. The main result is that if that coupling is small, then quantum fluctuations of the matrix background are small compared to the background, leading to a self-consistent semi-classical concept of spacetime and geometry.
Even though that result is hardly surprising, it has apparently not been spelled out up to now.
We provide a transparent derivation based on string modes, which generalizes immediately to rather generic noncommutative backgrounds.

This justifies the \newV{basic hypothesis and self-consistency of the } semi-classical approach to IKKT-type matrix models on $3+1$ dimensional backgrounds at weak coupling, and should remove lingering doubts.
In particular, the effective metric of such a background 
is then determined unambiguously through the kinetic term of the fluctuation modes \cite{Steinacker:2010rh}.
Most importantly, nontrivial matrix backgrounds provide a mechanism to get 3+1 dimensional (quantum) spacetime from a string-like theory without compactification.

In contrast to this weak coupling approach, one may of course follow a holographic strong-coupling approach leading to a different type of emergent geometry in 9+1 dimensions, cf. \cite{Ciceri:2025wpb,Komatsu:2024bop,Hartnoll:2024csr}. This is relevant in the deep quantum regime, hence it is much more difficult and requires analytical control over a strongly coupled theory. 

Finally, these statements do not apply to commutative backgrounds. A straightforward analysis of fluctuations on commutative backgrounds is clearly pathological, but it might be rescued by including nontrivial fermionic backgrounds, cf. \cite{Aoki:1998vn}.

We shall refrain here from discussing the  physical viability of noncommutative spacetime in the IKKT model;
 gravity on such backgrounds appears to be close to GR in intermediate regimes but distinct in more extreme scales, notably also in the IR \cite{Steinacker:2026qzk}. Holography on the other hand might be closer to vanilla GR but a priori lives in the wrong dimensions, and it seems hard to conceive how full-fledged 3+1-dimensional physics may arise in this way. Time will tell which of these approaches -- if any -- turns out to be successful.

\subsection*{Acknowledgments}

I am grateful for many useful discussions over the years on these and related topics, including D. Berenstein,  R. Brandenberger,  C-S Chu, S. Hartnoll, P-M. Ho, J. Karczmarek, H. Kawai,  S. Komatsu  A. Manta, J. Nishimura and  T. Tran, among many others, which motivated this paper. 
Comments are welcome!  

This work is supported by the Austrian Science Fund (FWF)
grant P36479.

\bibliographystyle{JHEP-2}
\bibliography{twistor}

\end{document}